\documentclass[twocolumn,prl,aps,floats,showpacs,amssymb]{revtex4}
\usepackage{graphicx}
\usepackage{amsmath}

\newcommand{\var}{{\rm var}\,}

\begin{document}

\title{Nonexponential decoherence and momentum subdiffusion\\ in
 a quantum L\'evy kicked rotator}

\author{Henning Schomerus$^{1}$  and Eric Lutz$^{2}$}
\affiliation{{}$^{1}$Physics Department,  Lancaster University,
Lancaster, LA1 4YB, UK\\ {}$^{2}$Department of Physics, University
of Augsburg, D-86135 Augsburg, Germany}

\date{\today}

\begin{abstract}
We investigate decoherence in the quantum kicked rotator
(modelling cold atoms in a pulsed optical field) subjected to
noise with power-law tail waiting-time distributions of variable
exponent (L\'evy noise). We demonstrate the existence of a regime
of nonexponential decoherence where the notion of a decoherence
rate is ill-defined. In this regime, dynamical localization is
never fully destroyed, indicating that the dynamics of the quantum
system never reaches the classical limit. We show that this leads
to quantum subdiffusion of the momentum, which should be
observable in an experiment.
\end{abstract}
\pacs{
03.65.Yz, %Decoherence; open systems%
05.40.Fb, %Random walks and Levy flights%
72.15.Rn %Localization effects%
} \maketitle

The transition  between quantum and classical dynamics is not yet
fully understood and the boundary between the two remains fuzzy.
Deep insight into the emergence of classical properties from
quantum mechanics is provided by the theory of decoherence
\cite{zur03,giu96}. In this framework, the unavoidable coupling to
an external environment induces a dynamical suppression of
interference effects between states of a quantum system, and thus
classical behavior. The disappearance of quantum interference can
often be described by a decoherence factor of the form ${\cal
D}(t) = \exp[- m \kappa\, (\Delta x)^2 \, t]$, where $m$ is the
mass of the particle, $\kappa$ the coupling strength and $\Delta
x$ the spatial separation of the interfering states. The
functional form of the decoherence factor, in particular its
exponential time dependence, has been verified in many experiments
\cite{bru96,mya00,amm98,kla98}. The boundary between the quantum
and the classical world can accordingly be explored by properly
tuning the value of the coherence time, $t_{c} = 1/ [m \kappa
\,(\Delta x)^2]$. This has been performed in various ways, either
by changing the separation $\Delta x$ as in the QED cavity and ion
trap experiments by the groups of Haroche \cite{bru96} and
Wineland \cite{mya00}, or by modifying the value of the coupling
constant $\kappa$ as done in the atom-optics experiments by the
groups of Christensen \cite{amm98} and Raizen \cite{kla98}. In yet
another approach, the quantum-classical border has been
investigated by increasing the mass $m$ of the system  as in the
double-slit experiments  by the group of Zeilinger \cite{hac03}.

Common to all these studies is the exponential functional
dependence of the decoherence factor. Our aim in this paper is to
widen our understanding of the quantum-to-classical crossover
 by proposing an environmental
coupling scheme for the atom-optical experiments
\cite{amm98,kla98} which allows the {\it controlled} tuning of the
functional form of ${\cal D}(t)$ itself, in particular of its time
dependence, and not just of its parameters. In this way, we are
able to uncover some hidden aspects of decoherence, especially of
its interplay with complex quantum dynamics
\cite{echo}.

A paradigm of complex quantum dynamics are cold atoms  exposed to
pulsed optical fields \cite{moo95}. For chaotic conditions the
classical motion in these systems exhibits momentum diffusion
which however is suppressed in the quantum regime because of
dynamical localization \cite{fis82}. In the recent decoherence
experiments \cite{amm98,kla98}, the quantum-classical transition
has been investigated by varying the amplitude of noisy pulses
acting on the atoms [coupling strength $\kappa$ in ${\cal D}(t)$].
This then results in a non--vanishing quantum momentum diffusion
with a renormalized diffusion constant. By contrast, we here
propose to modify the length $\tau$ of the time intervals between
the noisy pulses  \cite{osk03}. Specifically, we shall consider
that the intervals are generated in a renewal process with a
waiting-time distribution $w(\tau)$ that asymptotically behaves as
a power law, $w(\tau)\propto \tau^{-1-\alpha}$. Power-law tail
waiting-time distributions of this kind -- generally referred to
as L\'evy statistics \cite{shl93} -- occur naturally in many
physical systems: Two prominent examples being subrecoil laser
cooling \cite{sau99} and fluorescent intermittency of nanocrystal
quantum dots \cite{bro03} (in both these systems, $\alpha$ is
close to one half).  In the atom-optical setting, the statistics
are under the control of the experimentalist. By tuning the value
of the exponent $\alpha$, an additional parameter not present in
former decoherence studies, one is able to control the duration
between successive random kicks given by the mean waiting time,
$\overline\tau=\int_0^\infty d\tau \,\tau \,w(\tau)$. This offers
a unique opportunity to smoothly interpolate between a fully
coupled situation with many kicks (large $\alpha$) to an almost
isolated situation with very few kicks (small $\alpha$). L\'evy
kicked systems have lately been  investigated  at   the classical
\cite{bar03} as well as the quantum--mechanical level
\cite{sch05} and unusual  properties such as nonexponential
behavior and aging have been found.

Interestingly, for $\alpha \leq 1$, $\overline\tau$ becomes
mathematically infinite. We show that this divergence of the mean
waiting time dramatically affects the loss of phase-coherence of
the atoms: the decoherence factor stops being exponential in time
(instead it  is given by a Mittag--Leffler function) and the
concept of a decoherence rate becomes ill-defined. This results in
quantum  subdiffusion of the  momentum of the L\'evy kicked
rotator, a unique signature which would be directly accessible in
experiment.

{\it Model.}  The motion of an atom in a pulsed optical field can
be modelled as a kicked rotator, a connection that has been
established extensively in both theoretical and experimental works
\cite{chi78,izr90,moo95}. The kicked rotator can be simply seen as
a particle moving on a ring and periodically kicked in time. In
conveniently scaled coordinates, the Hamiltonian is given by
\begin{equation}
H=\frac{p^2}{2}+\sum_{n=-\infty}^\infty
K_n\cos\theta\,\delta(t-n+0^+),
\end{equation}
where the kicking potential depends on the $2\pi$-periodic
rotation angle $\theta$. We assume that the particle is initially
prepared in the zero-momentum state $\psi(0)=|p=0\rangle$. The
subsequent stroboscopic dynamics $\psi(t+1)=F(K_t)\psi(t)$ are
generated by the Floquet operator
\begin{equation}
F(K_t)=\exp(-i\hbar^{-1}\hat p^2/2)\exp(-i\hbar^{-1}K_{t}\cos\hat
\theta).
\end{equation}
This operator describes a  kick in which momentum changes by
$K_t\sin\theta$, followed by a free rotation in which the rotation
angle $\theta$ increases by $p$.

For constant (noiseless) kicks of strength $K_n=K\gtrsim 5$, the
{\em classical} dynamics is chaotic, and on average diffusive in
momentum direction: $\var p (t) \simeq D_{\rm cl} t$, where
$D_{\rm cl}\simeq  K^2/2$. In the quantum regime, the kicked
rotator exhibits dynamical localization \cite{fis82}, an
interference phenomenon which manifests itself in an exponentially
decaying envelope of the quasienergy eigenstates of $F(k)$ in the
momentum representation. As a result of  dynamical localization,
momentum diffusion is suppressed and the variance
\begin{equation}
\var p_0(t) \simeq D^* t^*[1-\exp(-t/t^*)]
\label{eq:varp0}\hspace{.25cm} (\mbox{with } D^*\simeq D_{\rm cl})
\end{equation}
saturates after the quantum break time $t^* \simeq D^*/ \hbar^2$
\cite{izr90}.

The dramatic difference between  classical and quantum momentum
dynamics makes the kicked rotator ideally suited for the
investigation of the crossover between both regimes. The crossover
can be induced by subjecting the particle to additional random
kicks --  simulating in such a way the noisy coupling to an
external environment \cite{amm98,kla98}.  In a noisy kick, the
kicking strength is slightly perturbed away from the mean value
$K$, $K_n=K+k_n$, where the perturbations $k_n$ are  random
numbers with average $\overline{k_n}=0$ and variance
$\overline{k_n^2}=\kappa$. The parameter $\kappa$ defines  the
strength of the noise. The time intervals between subsequent kicks
are generated by a renewal process with waiting distribution
$\omega(\tau)$. Unlike the familiar  Gaussian white noise, L\'evy
noise with a power--law waiting time distribution is a
nonstationary process with an autocorrelation function that
depends explicitly on the two time variables.

The effect of  noise on the quantum kicked rotator is best
understood using the formalism developed by Ott, Antonsen and
Hanson \cite{ott84}. In the absence of noise the quantum system is
fully coherent: the quasienergy eigenstates of the Floquet
operator $F(K)$ are exponentially localized in momentum space,
chaotic diffusion is strongly suppressed and the momentum
diffusion constant asymptotically vanishes. The effect of the
external noise is to couple the quasienergy states and to induce
transitions between them. As a result, dynamical localization is
extenuated and quantum diffusion takes place. There is thus an
intimate connection between quantum momentum diffusion and  phase
coherence in  the kicked rotator.

A systematic, perturbative approach for calculating the quantum
diffusion constant  for weak stationary noise with arbitrary
correlations has been put forward by Cohen \cite{coh91}. In  this
case, the survival probability of the quasienergy eigenstates  is
found to decay  exponentially over time,  ${\cal
D}(t)=\exp(-t/t_c)$. In the limit of weak noise, $\kappa\ll
K^4/\hbar^4$,  the coherence time is given by  $t_c\simeq
2\hbar^2/\kappa$. The resulting momentum spreading is accordingly
\cite{coh91}
\begin{equation}
\var p\approx \frac{D^*}{1+t_c/t^*}t+
\frac{D^*t^*}{(1+t^*/t_c)^2}[1-\exp(-t/t^*-t/t_c)].
\label{eq:crossover}
\end{equation}
Hence, for stationary noise when  quantum coherence is lost
exponentially, the momentum diffuses asymptotically with a
renormalized diffusion constant $D^*/(1+t_c/t^*)$. Equation
(\ref{eq:crossover}) has been successfully applied to quantify the
decoherence process in the experiment \cite{kla98}.

{\it Results.}   In the following, we extend the calculation of
the quantum diffusion constant beyond the perturbative regime: for
the present case of uncorrelated perturbations $\kappa_n$, a
random--phase approximation allows  to study  decoherence induced
by  {\it strong nonstationary} noise (details of the derivation
will be presented at the end of this paper). In a general renewal
process, the number of noisy events $N(t',t'')$ within an interval
$[t',t'']$ (technically known as the {\em inverse time} of the
process)  is a statistical quantity. Within the random-phase
approximation, the survival probability in a quasienergy
eigenstate over this time interval is simply given by the
decoherence factor
\begin{equation} {\cal
D}(t',t'')=\overline{\exp\left[- N(t',t'')/t_{c}\right]},
\label{eq:varpB}
\end{equation} where $t_{c}$ is the
coherence time of the corresponding stationary noise process with
the same noise intensity. The variance of the momentum can then be
obtained from the following expression (see details  below):
\begin{eqnarray}
\label{eq:varp} &&\var p(t)=\var p_0(t)\,{\cal
D}(t,0)+\frac{\kappa}{2}\overline{N(t,0)}\nonumber\\
&& +\int_0^t ds\,\var p_0(t-s)\,\partial_s {\cal
D}(t,s)\nonumber\\
&& -\int_0^t ds\,\var p_0(s)\,\partial_s {\cal D}(s,0)
\nonumber\\
&&-\int_0^t ds'\int_0^{s'} ds''\,\var p_0(s'-s'')\,\partial_{s'}
\partial_{s''}{\cal
D}(s',s'')  .
\end{eqnarray}
Here $\var p_0(t)$ is the variance of the momentum in absence of
the noise. The second term on the right-hand side becomes
important for large noise strength.

Stationary  noise can be regarded as a special case of a renewal
process with $w(\tau)=\delta_{1,\tau}$. In the latter situation,
the number of noisy events is simply $N(t',t'') = |t'-t''|$ and
the decoherence function  decays exponentially, ${\cal
D}(t',t'')=\exp\left(-|t'-t''|/t_{c}\right)$. Equation
(\ref{eq:varp}) then provides a generalization of the perturbative
result (\ref{eq:crossover}) which accounts exactly for all
correlations of the noiseless dynamics. Expression
(\ref{eq:crossover}) is recovered when the noiseless dynamics is
approximated by the variance (\ref{eq:varp0}).

We now evaluate the decoherence factor (\ref{eq:varpB}) for a
renewal process with a general waiting time distribution. The
Laplace transform of ${\cal D}(t,0)$ with respect to $t$ reads
\begin{equation}
{\cal D}(u,0)=\frac{1}{u(1+f(u)/t_c)},
\end{equation}
where $f(u)=w(u)/(1-w(u))$ is the Laplace transform of the {\it
sprinkling distribution} $f(t)=\partial_t\overline{N(t,t')}$
[$f(t)$ is simply the probability that there is a noise event at
time $t$]. For two time arguments, we have  (see also
Ref.~\cite{sch05} for similar results),
\begin{equation}
{\cal D}(t',t'')={\cal D}(t',0)+t_c^{-1}\int_0^{t''}ds\,f(s){\cal
D}(t'-s,0).
\end{equation}

\begin{figure}[t]
\includegraphics[width=\columnwidth]{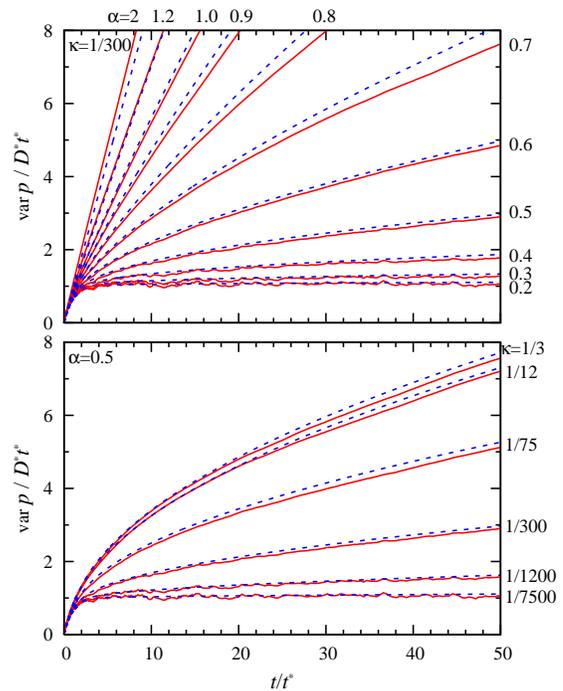}
\caption{(color online). Time dependence of the momentum spreading
$\var p(t)$ for kicked rotators  which are subjected to L{\'e}vy
noise generated by a Yule-Simon distribution. The  results of
numerical simulations (solid curves) are compared to the
theoretical predictions from Eq.\ (\ref{eq:varp}) (dashed curves).
In the upper panel, the exponent $\alpha$ varies while the noise
strength $\kappa=1/300$ is fixed. The bottom panel shows results
for fixed $\alpha=0.5$ and varying noise strength $\kappa$.
\label{fig:1}}
\end{figure}

Two cases have to be distinguished: For L\'evy noise with an
exponent $\alpha>1$, the sprinkling distribution approaches a
constant for long times, $f(t)\simeq \overline{\tau}^{-1}$,  and
the decoherence function ${\cal D}(t',t'') \simeq
\exp[-|t'-t''|/(\overline\tau t_{c})]$ has still an asymptotic
exponential form. We thus have the important result that the
effective coherence time $\overline\tau t_{c}$ is directly
proportional to the mean waiting time of the noisy kicks and
therefore increases with decreasing exponent $\alpha$. In
particular, the effective coherence time becomes {\it infinitely}
large when $\alpha$ drops below unity, indicating that the
classical limit cannot be attained in finite times. Clearly, the
functional form of the decoherence factor has to change as well.
In order to determine  ${\cal D}(t',t'')$ for   L\'evy noise  with
$\alpha<1$, we use the mathematical equivalence of Eq.\
(\ref{eq:varpB}) to the moment-generating function
$M(\xi)=\overline{\exp(\xi N)}$ of the inverse time $N$, evaluated
at $\xi=-1/t_{c}$ \cite{MLpaper}. We then find
\begin{equation}
{\cal D}(t,0)\approx E_\alpha\left(-
\Gamma(1+\alpha)\overline{N(t,0)}/t_c \right), \label{eq:MLp}
\end{equation}
where $E_\alpha(z)=\sum_{n=0}^\infty z^n/\Gamma(\alpha n+1)$ is
the Mittag-Leffler function \cite{MLfunction}. Asymptotically,
$N(t,0)\sim t^{\alpha}\sin(\pi\alpha)/(\pi c)$. The initial decay
of the decoherence function is therefore  a stretched exponential,
${\cal D}(t,0) \simeq \exp\{t^\alpha /[\Gamma(-\alpha) c t_c]\}$,
while for large times it crosses over to the power law  ${\cal
D}(t,0)\simeq (c t_c/\alpha) t^{-\alpha}$. The functional
dependence of the decoherence factor can thus be changed in a
controlled manner by tuning the value of the exponent $\alpha$.
The consequences of such a modification can be directly detected
in the momentum diffusion of the kicked rotator: As $\alpha$
approaches unity from above, the decoherence time increases, and
the asymptotic quantum diffusion coefficient $D^*/(1+\overline\tau
t_c/t^*)$ vanishes at $\alpha=1$. For $\alpha\leq 1$ it follows
from Eq.\ (\ref{eq:varp}) that the momentum spreads
subdiffusively,
\begin{equation}
\var p(t)  \simeq \frac{\var p_0(\infty)}{t_c}\,\overline{N(t,0)}\simeq
\frac{D^*t^*}{t_c}\frac{\sin\pi\alpha}{\pi c} t^{\alpha}.
\label{eq:subdiff}
\end{equation}

The  full time dependence of the momentum spreading in
 kicked rotators with specific implementations of L\'evy  noise is shown in Fig.\
\ref{fig:1}. The regular kicking strength is set to $K=7.5$, while
$\hbar = 2\pi\times 577/13872$.
 The random numbers
$k_n$ are taken from a uniform box distribution over an interval
$(-W,W)$, so that $\kappa=W^2/3$. The waiting-time distribution is
of the Yule-Simon form $\omega (\tau) = \alpha
\,\Gamma(\tau)\Gamma(\alpha +1)/\Gamma(\tau +\alpha+1)$, with
asymptotic behavior $\omega(\tau)\sim \alpha\,
\Gamma(\alpha+1)/\tau^{\alpha+1}$ and a mean waiting time
$\overline\tau=\alpha/(\alpha-1)$ \cite{sim55}.
 In the figure, the solid lines are computed by
extensive numerical simulations of the dynamics of kicked rotators
over $10^5$ time steps, averaged over 100-1000 realizations of the
noise. The dashed lines are evaluated from Eq.\ (\ref{eq:varp}),
where $\var p_0(t)$ is approximated by Eq.\ (\ref{eq:varp0}) and
$D^*=\hbar^2t^*=45.28$ is obtained by a single-parameter fit to
the results of the numerical simulations without noise. The
decoherence function (\ref{eq:varpB}) for the Yule-Simon
distribution is determined setting $t_c=2\hbar^2/\kappa$. We find
 good agreement between theory and numerics, both as a
function of the exponent $\alpha$ as well as a function of the
noise strength $\kappa$.  In particular, the numerics confirm the
subdiffusive long-time dynamics (\ref{eq:subdiff}) for $\alpha\leq
1$, as well as the diffusive behavior for $\alpha>1$.

{\it Details of the calculation}. The derivation of our main
result (\ref{eq:varp}) along with Eq.\ (\ref{eq:varpB}) is based
on the observation that the phase of the transition amplitudes
$\langle r | F(K_l)| s\rangle\equiv F_{r,s}(K_l)$ between
quasienergy eigenstates $|r\rangle$ and $|s\rangle$ depends
strongly on the random detuning $k_l=K_l-K$ of the kick. In the
presence of noise we  set
\begin{equation}
\overline {F_{r,s}(K_l)}=\delta_{r,s} \exp[i\omega_r-1/(2t_{c})]
\label{eq:rpa}
\end{equation}
while in the absence of noise $
F_{r,s}(K)=\delta_{r,s}\exp(i\omega_r)$, where $\omega_r$ is the
quasienergy of the state.

The transition amplitudes $F_{r,s}$ enter the variance of the
momentum $ \var p(t)=\sum_{t',t''=0}^{t-1} C(t',t'')$ when the
force-force correlation $C(t',t'')=\overline{\langle
K_{t'}K_{t''}\sin\theta_{t'}\sin\theta_{t''}\rangle}$ is expanded
in the basis of quasienergy eigenstates \cite{coh91},
%\begin{widetext}
\begin{eqnarray}
&&C(t',t'')=\sum_{\{r_n,s_n\}} \langle
 r_{t'} | \sin\theta| s_{t'}  \rangle  \langle
s_{t''} | \sin\theta| r_{t''}  \rangle
\times{}
\nonumber\\
&&\times{}
\quad \prod_{l,m=t''}^{t'-1}
\overline{K_{t'}K_{t''}F^*_{r_{l+1},r_{l}} (K_l)
F_{s_{m+1},s_{m}}(K_m)}. \label{eq:c}
\end{eqnarray}
%\end{widetext}
In the sum (\ref{eq:c}),  only those terms with satisfy either
 (a) ${r_n\equiv r}$ and ${s_n\equiv s}$, or (b) $r_n\equiv
s_n$ survive the average in the random-phase approximation
(\ref{eq:rpa}). The  terms (b), however,  have a zero contribution
since they  are multiplied by  $\langle r| \sin \theta |r
\rangle=0$,  a property which  follows from the symmetry
$(\theta,p)\to (-\theta,-p)$ of the kicked rotator. Each of the
terms (a) further carries a weight factor $\exp[-N(t',t'')/t_c]$,
which has to be averaged using Eq.\ (\ref{eq:rpa}). The averaging
procedure over the renewal process then leads to the decoherence
factor ${\cal D}(t',t'')$ defined in Eq.\ (\ref{eq:varpB}).
Moreover, for $t'=t''$ the noise also gives rise to a contribution
to the typical force acting on the particle of the form
$\overline{K_{t'}^2}\langle \sin\theta^2\rangle=K^2/2 +(\kappa/2)
f(t')$.
 Collecting all terms, this yields
\begin{equation}
C(t',t'')=C_0(t',t''){\cal
D}(t',t'')+\frac{\kappa}{2}f(t')\delta_{t',t''},
\end{equation}
where  $C_0(t',t'')$ is the force-force correlation function in
the absence of  noise. Equation (\ref{eq:varp}) then follows by
converting  sums into integrals and integrating by parts.

{\it Conclusions.} Nonexponential  relaxation is familiar from the
physics of complex systems -- such as disordered crystals and
glassy materials -- and occurs whenever there is no clear
separation of time scales between  system and  environment
\cite{ric94}. At the quantum level, the absence of  a  scale
separation manifests itself in a nonexponential  loss of phase
coherence. We have shown that such a behavior can be induced in
atom-optical systems by properly engineering the  environment. By
applying L\'evy noise with a controllable exponent $\alpha$, the
environment time scale --  the mean waiting time $\overline{\tau}$
between noisy events --  is finite for $\alpha>1$ and can  be made
divergent by choosing $\alpha<1$. For the quantum  L\'evy kicked
rotator, the power--law decay of the decoherence factor  and the
nonexistence of a finite decoherence rate reveals itself in
quantum subdiffusion of the momentum, a distinct signature that
can easily be observed experimentally using cold atoms in pulsed
optical fields. The high degree of control and tunability of
atom-optical systems make the L\'evy kicked rotator  an ideal tool
for the study of the crossover from fast to slow decoherence.
Complex systems with algebraic relaxation are known to never
attain equilibrium; the possibility of having noisy quantum
systems that never become classical  appears as a fascinating
prospect.

This work was supported by the European Commission, Marie Curie
Excellence Grant MEXT-CT-2005-023778 (Nanoelectrophotonics) and
the Emmy Noether Program of the DFG under contract LU1382/1-1.

\end{document}